\documentclass[journal=jacsat,manuscript=article]{achemso}

\usepackage[version=3]{mhchem} 
\usepackage{comment}
\usepackage{xr}
\usepackage{hyperref}
\DeclareUnicodeCharacter{2009}{\,} 

\newcommand*\mycommand[1]{\texttt{\emph{#1}}}

\author{Mohammad Amin Akhound}
\affiliation{CAMD, Department of
Physics, Technical University of Denmark, DK - 2800 Kongens Lyngby,
Denmark}
\email{akhound@dtu.dk}
\author{Karsten Wedel Jacobsen}
\affiliation{CAMD, Department of
Physics, Technical University of Denmark, DK - 2800 Kongens Lyngby,
Denmark}
\author{Kristian Sommer Thygesen}
\affiliation{CAMD, Department of
Physics, Technical University of Denmark, DK - 2800 Kongens Lyngby,
Denmark}
\email{thygesen@fysik.dtu.dk}

\title[An \textsf{achemso} demo]
  {Activating the Basal Plane of 2D Transition Metal Dichalcogenides via High-Entropy Alloying}

\begin{document}


\begin{tocentry}

\centering
\includegraphics[height=1.8cm,keepaspectratio]{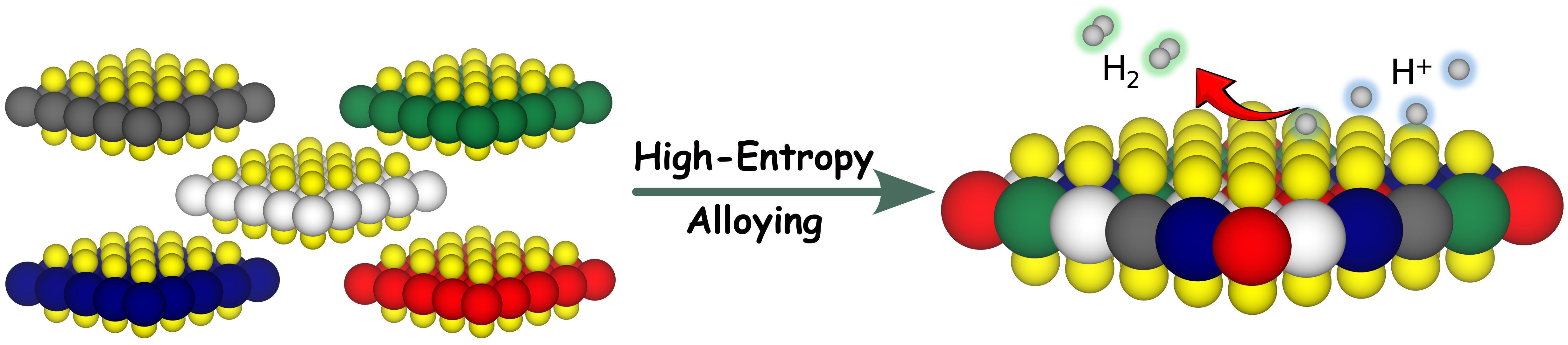}

\end{tocentry}

\begin{abstract}
 Two-dimensional (2D) materials, such as transition metal dichalcogenides (TMDCs) in the 2H or 1T crystal phases, are promising (electro)catalyst candidates due to their high surface to volume ratio and composition of low-cost, abundant elements. While the edges of elemental TMDC nanoparticles, such as MoS$_2$, can show significant catalytic activity, the basal plane of the pristine materials are notoriously inert, which limits their normalized activity. Here we show that high densities of catalytically active sites can be formed on the TMDC basal plane by alloying elements that prefer the 2H (1T) phase into a 1T (2H) structure. The global stability of the alloy, in particular whether it crystallizes in the 2H or 1T phase, can be controlled by ensuring a majority of elements preferring the target phase.  We further show that the mixing entropy plays a decisive role for stabilizing the alloy implying that high-entropy alloying becomes essential. Our calculations point to a number of interesting non-precious hydrogen evolution catalysts, including (CrHfTaVZr)S$_2$ and (CrNbTiVZr)S$_2$ in the T-phase and (MoNbTaVTi)S$_2$ in the H-phase. Our work opens new directions for designing catalytic sites via high-entropy alloy stabilization of locally unstable structures.
\end{abstract}

\section{Introduction}

In the quest for sustainable energy solutions, hydrogen (H$_2$) will undoubtedly play a major role. The hydrogen evolution reaction (HER), occurring at the cathode during electrocatalytic water splitting, plays a crucial role in converting electrical energy into hydrogen gas\cite{Seh2017CombiningDesign}. Noble metals such as platinum \cite{Zhang2023Superaerophilic/superaerophobicTransfer}, ruthenium\cite{Chen2023RevealingReaction}, and palladium \cite{Liang2024ContinuousLevel} are excellent HER catalysts; however, their high costs and scarcity presents a major barrier to their large-scale application. Consequently, there is a growing focus on developing cost-effective and efficient HER catalysts based on non-precious metals. Recently, a wide range of alternatives including transition metal dichalcogenides (TMDCs) \cite{Guo2022ChargeActivity, Sun2022AtomicLevelModelling}, phosphides \cite{Banerjee2023RelationshipReaction, Bodhankar2022NanostructuredReview}, carbides \cite{Yu2022FacileReaction, Tran2024ReviewRecentReaction}, and nitrides \cite{Su2022InsightsReaction, Jin2021StableNitride}, have been identified as promising HER catalysts. Among these materials, TMDCs have been extensively investigated due to their layered structures and associated high surface to volume area \cite{VanNguyen2023ElectrocatalystsOverview}. Unfortunately, the layered structure that allows the individual two-dimensional (2D) sheets to be easily separated (thereby enhancing their surface area) is a direct consequence of the chemical inertness of the layers. Therefore, the catalytic activity of the TMDC nanoparticles is primarily due to highly reactive edge sites while the basal planes exhibit relatively lower activity \cite{Jaramillo2007IdentificationNanocatalysts, Pakhira2023RevealingReaction}. Although multiple strategies, including doping \cite{Pattengale2020DynamicEvolution, Wu2023DopantHydrogen}, defect engineering \cite{Wang2023VacancyConfiguration, Su2023DefectengineeredReaction}, intercalation\cite{Americo2024EnhancingSelf-Intercalation}, and strain manipulation \cite{Liu2022StrainReaction, Voiry2013EnhancedEvolution}, have been explored as a means to enhance the basal plane reactivity, there is still substantial room for improvement.

High-entropy alloys (HEAs) have emerged as an interesting class of materials with promising applications in the field of energy storage and electrocatalysis \cite{Miracle2017AConcepts, Wu2024HighEntropyDevelopment}. HEAs are single-phase crystals composed of five or more principal elements in nearly equal atomic ratios, resulting in a high configurational entropy that stabilizes the structure. \cite{Liu2022RecentAlloys, Zhang2021TheReaction}. The possibility to tune composition and concentrations allows for detailed control of physical and chemical properties, including optimization of reaction kinetics and catalytic performance \cite{Batchelor2019High-EntropyElectrocatalysis, Wang2022Two-DimensionalReaction, Chen2024UnusualReactions}. Many HEAs have good mechanical properties, such as enhanced hardness, superior elongation capacity, and high thermal stability, which is important to ensure long-term performance of the catalyst under harsh electrochemical conditions. \cite{Ren2023High-entropyApplications}  Finally, the so-called cocktail effect, referring to synergistic interactions among multiple elements, can lead to emergent properties that surpass those of the individual elements, leading to superior performance, e.g. for catalysing complex chemical reactions \cite{Zhang2023High-entropyChallenges}. 

HEAs based on two-dimensional (2D) materials promise to combine the compositional tunability of HEAs with the structural flexibility and large surface to volume ratio of 2D materials. Recently, a number of TMDC 2D HEAs have been successfully synthesized\cite{Ying2021High-EntropyTrisulfidesb,Qu2023AElectrocatalysis,Cavin20212DElectrocatalysis}. In particular, (MoWReMnCr)S$_2$\cite{Qu2023AElectrocatalysis} and (MoWVNbTa)S$_2$ \cite{Cavin20212DElectrocatalysis} were shown to possess exceptional thermal stability and decent electrocatalytic activity for HER and CO$_2$ reduction reaction, respectively. Notably, both of these TMDC compounds adopt the 2H crystal phase, which is also the most stable phase of the individual constituent elements. Density functional theory (DFT) calculations indicate the potential stability of TMDC alloys with various compositions in both the 1T and 2H phases\cite{Deshpande2022EntropyStudy}. However, a systematic exploration of the stability and catalytic performance of HEA TMDCs as well as microscopic understanding of the relation between these two key properties is still missing. 

In this work, we develop a general strategy for designing stable catalysts with highly active sites, and use it to discover several TMDC HEAs with basal plane reactivities optimised for the HER. The design strategy is based on the observation that thermodynamic stability correlates with surface reactivity: Less stable compounds bind adsorbates (e.g. hydrogen) stronger and vice versa. This suggests that good catalysts are always a result of a stability/activity compromise. However, we then show that this trend also holds locally. This implies that by alloying locally unstable elements into a globally stable structure it is possible to combine good overall stability with high local reactivity. We first establish the design principle on basis of an analysis of 18 unary and 72 binary TMDCs in both the 1T and 2H phases. We then demonstrate that catalytic reactivity can be enhanced by the integration of metal elements that prefer the 2H phase into the 1T phase (or vice versa). Importantly, high entropy alloying is essential to stabilise the target phase. From our systematic study of more than 400 TMDCs, including ternary, quaternary, and pentanary alloys, we observe a clear trend: increasing the number of constituent elements enhances both stability and catalytic performance. Detailed calculations of mixing enthalpy, free energy, and hydrogen adsorption energy then identifies 20 stable pentanary alloys, 10 of which exhibit hydrogen adsorption energies within the optimal range for the HER. All the results of our work is available in the open database: https://cmrdb.fysik.dtu.dk/2dalloys 

\section{Methods}

\subsection{Density Functional Theory Calculations}
All DFT calculations were performed using the projector augmented wave (PAW) method \cite{Blochl1994ProjectorMethod} as implemented in the GPAW electronic structure code \cite{Enkovaara2010ElectronicMethod,Mortensen2024GPAW:Calculations} with the Perdew-Burke-Ernzerhof (PBE) exchange-correlation functional \cite{Perdew1996GeneralizedSimple}. Spin polarization was consistently used in all calculations with a Fermi smearing of 0.05 eV and plane-wave cutoff of 800 eV. For unary, binary, ternary and quaternary alloys a $k$-point density of 6 \AA$^{-1}$ while a coarser mesh of density of 3 \AA$^{-1}$ was used for pentanary alloys.  A vacuum spacing of at least 20 \AA\ was used along the $z$ direction to ensure adequate separation between periodic layers. Both the unit cell and the atomic positions were relaxed until the forces were below 0.01 eV/\AA. The DFT calculations were managed using the MyQueue \cite{Mortensen2020MyQueue:System} task scheduler.

The choice of the PBE functional is based on its well-established balance between accuracy and computational efficiency. Previous studies have demonstrated that hydrogen adsorption energy is not significantly impacted by the choice of functional \cite{Rangarajan2023A2H-MoS2, Castelli2018EffectsMgO100}. To further validate this, we compared the PBE with two other functionals for the type of systems considered in this study. As illustrated in Figure S1, the H adsorption energies on the unary TMDCs calculated using PBE, RPBE\cite{Hammer1999ImprovedFunctionals}, and BEEF-vdW\cite{Wellendorff2012DensityEstimation} are in very close agreement, thus justifying our choice of the PBE. 




\subsection{Atomic Structures}
The atomic structures were generated using the Atomic Simulation Environment (ASE) \cite{HjorthLarsen2017TheAtoms}. Both the P$\overline{3}$m1 (1T phase) and P$\overline{6}$m2 (2H phase) space groups were considered for all TMDCs in their 2D monolayer forms. All generated alloy structures contain an equimolar composition, ensuring equal concentrations of metal atoms across different alloys. For unary TMDCs and their binary alloys, 2$\times$2 supercells were considered (Figure S6). To preserve equimolar concentration in more complex systems, the supercell size was increased to 3$\times$3, 4$\times$4, and 5$\times$5 for ternary, quaternary, and pentanary alloys, respectively. Given the wide range of possible configurations for TMDC alloys, three different random configurations were generated for each phase of the ternary (Figure S7) and quaternary (Figure S8) alloys. The most stable configuration in each case was selected for further analysis and hydrogen adsorption studies. For the pentanary alloys, special quasi-random structures (SQSs) \cite{Zunger1990SpecialStructures} were generated using the alloy theoretic automated toolkit (ATAT) \cite{vandeWalle2002TheGuide} software package. This method simulates the chemical disorder characteristics of HEAs while maintaining a computationally feasible cell size. SQSs, comprising 75 atoms, were constructed for the 1T and 2H phases of pentanary alloys (Figure S9), ensuring a realistic representation of 2D HEAs.

\subsection{Free Energy Calculations}
The free energy of mixing, \( \Delta G_{\mathrm{mix}} \), at a given temperature, \( T \), is given by:

\[
\Delta G_{\mathrm{mix}} = \Delta H_f - T \Delta S_{\mathrm{mix}}
\]

In this equation, \( \Delta H_f \) is the formation enthalpy of the alloy, which is determined using DFT calculations, and \( \Delta S_{\mathrm{mix}} \) represents the entropy of mixing. The formation enthalpy is calculated as the difference between the total energy of the alloy and the weighted average of the constituent elements, as given by:

\[
\Delta H_f = E^{\mathrm{Alloy}} - \sum_{i} n_i E_i^{\mathrm{phase}},
\]
where \( E^{\mathrm{Alloy}} \) is the total energy of the alloy normalized by the cell size to obtain the energy per metal atom, \( n_i \) is the molar fraction of the \( i \)-th component, and \( E_i^{\mathrm{phase}} \) refers to the energy per metal atom of the unary TMDCs in their most stable phases.

The mixing entropy (or configurational entropy), \( \Delta S_{\mathrm{mix}} \), is calculated from the expression
\[
\Delta S_{\mathrm{mix}} = -R \sum_{i=1}^{N} c_i \ln c_i,
\]
where \( c_i \) represents the molar concentration of the \( i \)-th component and \( R \) is the universal gas constant. For an equimolar alloy with \( N \) components, this simplifies to:

\[
\Delta S_{\mathrm{mix}} = R \ln N
\]

Negative values of \( \Delta G_{\mathrm{mix}} \) indicate a miscible alloy, while positive values suggest immiscibility, i.e. phase separation. All reported free energies are evaluated at $T=1000$ K, which is comparable to the temperatures employed in the synthesis of 2D HEAs\cite{Cavin20212DElectrocatalysis, Nemani2021High-EntropyTiVCrMoC3, Ying2021High-EntropyTrisulfides}. Alloys with \( \Delta G_{\mathrm{mix}} < 0.05 \, \text{eV} \) at 1000 K, were classified as stable/miscible and selected for hydrogen adsorption calculations.

In addition to the free energy of mixing, the energy above the convex hull ($\Delta \mathit{H}_{\mathrm{hull}}$) is a crucial metric to evaluate the thermal stability and synthesizability of the predicted alloys. The $\Delta \mathit{H}_{\mathrm{hull}}$ quantifies the energy difference between the formation energy of the alloy and the lowest energy combination of its constituent elements, considering all possible decomposition phases. Alloys with $\Delta \mathit{H}_{\mathrm{hull}}$ values close to zero are considered thermodynamically stable because they represent the most stable configuration of elements and are less likely to decompose into other phases. The $\Delta \mathit{H}_{\mathrm{hull}}$ values for all predicted HEAs were calculated relative to the alloys considered in this work and the entries in the Open Quantum Materials Database (OQMD).\cite{Saal2013MaterialsOQMD, Kirklin2015TheEnergies}

\subsection{Hydrogen Adsorption}
The free energy of hydrogen adsorption (\( \Delta G_{\mathrm{H}^*} \)) is determined as
\[
\Delta G_{\mathrm{H}^*} = \Delta \mathit{H}_{\mathrm{ads}}^{\mathrm{H}^*} + \Delta E_{\text{ZPE}} - T\Delta S_{\mathrm{H}^*}
\]
where \( \Delta \mathit{H}_{\mathrm{ads}}^{\mathrm{H}^*} \) represents the hydrogen adsorption energy,
\[
\Delta \mathit{H}_{\mathrm{ads}}^{\mathrm{H}^*} = \mathit{E}_{\mathrm{DFT}}\left({\mathrm{H}^*}\right) - \mathit{E}_{\mathrm{DFT}}\left({*}\right) - \frac{1}{2} \mathit{E}_{\mathrm{DFT}}\left({\mathrm{H}_2}\right), 
\]
and $\Delta E_{\text{ZPE}}$ is the difference in zero point energy between the adsorbed hydrogen and H$_2$ in the gas phase, 
\[
\Delta E_{\text{ZPE}} = \mathit{E}_{\mathrm{ZPE}}\left({\mathrm{H}^*}\right) - \frac{1}{2} \mathit{E}_{\mathrm{ZPE}}\left({\mathrm{H}_2}\right)
\]
The vibrational frequencies of the adsorbed hydrogen atom are relatively consistent across the TMDC surfaces, allowing us to approximate \( \Delta E_{\text{ZPE}} \approx 0.05 \)eV, based on values from prior studies \cite{Tang2016MechanismPrinciples, Ling2019EnhancingHeterostructuresb, Hinnemann2005BiomimeticEvolution, Chou2015UnderstandingDichalcogenide}.

The entropy contribution (\( \Delta S_{\mathrm{H}^*} \)) simplifies as 
\[
\Delta S_{\mathrm{H}^*} = S({\mathrm{H}^*}) - \frac{1}{2} S({\mathrm{H}_2})\approx -\frac{1}{2} S({\mathrm{H}_2})
\]
because the entropy of the adsorbed hydrogen atom is negligible. At standard conditions (298 K and 1 bar), \( S({\mathrm{H}_2}) \) is approximately 130 J·mol\(^{-1}\)·K\(^{-1}\) \cite{Atkins2006PhysicalChemistry}, resulting in an entropy contribution of roughly 0.2 eV. Thus, the free energy of hydrogen adsorption becomes
\[
\Delta G_{\mathrm{H}^*} = \Delta \mathit{H}_{\mathrm{ads}}^{\mathrm{H}^*} + 0.25 \, \text{eV}
\]
This approximation, which incorporates both zero-point energy and entropy corrections, is widely adopted and validated by several studies \cite{Tang2016MechanismPrinciples, Nrskov2005TrendsEvolution, Huang2025Polyacrylonitrile-basedEvolution, Zheng2014HydrogenElectrocatalyst}. Consequently, \( \Delta \mathit{H}_{\mathrm{ads}}^{\mathrm{H}^*} = -0.25 \, \text{eV} \), corresponding to \( \Delta G_{\mathrm{H}^*} = 0 \, \text{eV} \) has been widely accepted as the condition for the ideal HER catalyst surface. This value is highlighted by the gray dotted lines in several plots presented in this study.

\section{Results and discussion}

\subsection{Phase-Dependent Catalytic Activity in TMDCs}
We consider TMDCs containing sulphur and transition metals from groups IV, V, and VI (Ti, Zr, Hf, V, Nb, Ta, Cr, Mo, W). These transition metals (TMs) were chosen based on several considerations:

(a) The selected TMs have similar atomic radii and relatively uniform electronegativities, which is essential to maintain structural stability, avoid lattice distortions, and minimize charge imbalance upon alloying. These properties are essential to circumvent phase separation and ensure the formation of stable and homogeneous alloys.

(b) TMDCs based on group-IV transition metals crystallise in the 1T phase while group-VI metals crystallise in the 2H phase. In contrast, TMDCs made from group-V transition metals can exist in both the 1T and 2H phases. This means that the structural phase becomes a design parameter (in addition to the chemical composition and concentrations), in the search for stable and catalytically active HEAs.

(c) Group-VI TMDCs are known for their good catalytic activity, but their performance depends on their phase. The semiconducting 2H phase exhibits low basal plane activity (though active edges) whereas the metallic 1T phase provides enhanced activity\cite{voiry2013conducting} but lacks stability. The fact that the group-IV TMDCs are most stable in the 1T phase, suggests that incorporating these elements into a HEA together with group-VI elements, could provide a means to stabilize the latter in their catalytically active 1T phase. 

We first investigate the stability and catalytic activity (here quantified by the ability to bind hydrogen with $\Delta H_{\mathrm{ads}}=-0.25$eV) of the unary TMDCs. The structures and the energy above the convex hull of 18 unary TMDCs (nine in each phase) were extracted from the Computational 2D Materials Database (C2DB) \cite{Haastrup2018TheCrystals, Gjerding2021RecentC2DB}. The hydrogen atoms were adsorbed onto a basal plane sulfur atom, which is the most stable adsorption site.  As depicted in Figure 1a, group-VI TMDCs (Cr, Mo, W) are stable in the 2H phase ($\Delta \mathit{H}_{\mathrm{hull}} = 0$), while they exhibit significant instability in the 1T phase, with $\Delta \mathit{H}_{\mathrm{hull}}$ values ranging from 0.54 to 0.9 eV/metal atom. Despite their instability, the 1T phase exhibits significantly higher catalytic activity, as evidenced by lower hydrogen adsorption energies. A similar but inverse trend is observed for group-IV TMDCs (Ti, Zr, Hf) where the 1T phase is more stable and the 2H phase exhibits enhanced catalytic activity. Group-V TMDCs (V, Nb, Ta) display only a slight preference for the 2H phase with an energy difference between the 1T and 2H phases of less than 0.1 eV/TM atom. This minimal energy difference suggests that group-V TMDCs could be incorporated into both 1T and 2H phases without penalising the energy, but contributing to a higher configurational entropy. 

Our analysis of unary TMDCs suggests that less stable materials are more active towards the HER. To further substantiate this claim, we analyze the hydrogen adsorption on all possible binary alloys formed from the nine transition metals. As shown in Figure 1b, the trends observed for unary TMDCs are also evident in the binary alloys. Specifically, the structures on the left side of the plot, which are more stable in the 1T phase, exhibit stronger hydrogen adsorption in the unstable 2H phase. Conversely, structures on the right side, which are more stable in the 2H phase, exhibit stronger hydrogen adsorption in the unstable 1T phase. Additionally, the data indicate that alloys with a significant difference in phase stability (i.e. those further away from the center of the $x$-axis) show markedly higher catalytic activity in the less stable phase. In contrast, alloys located closer to the center of the $x$-axis, where the stability difference between phases is minimal, generally exhibit similar hydrogen adsorption energies in the 2H and 1T phases.

\begin{figure}[h]
\centering
  \includegraphics[height=6cm]{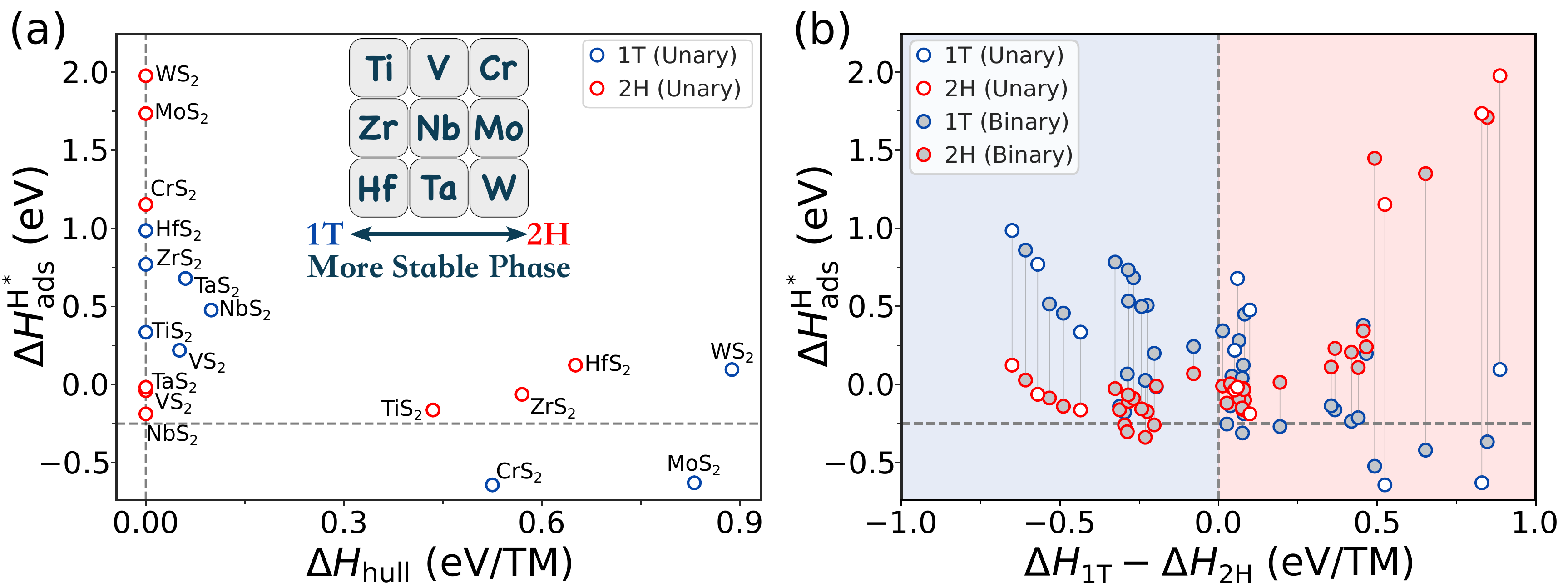}
  \caption{(a) Correlation between hydrogen adsorption energy ($\Delta \mathit{H}_{\mathrm{ads}}^{\mathrm{H}^*}$) and energy above convex hull ($\Delta \mathit{H}_{\mathrm{hull}}$) of unary TMDCs in the 1T and 2H phases. (b) Catalytic activity of unary and binary TMDCs, relative to their more stable phase. Structures on the left side are more stable in the 1T phase, while those on the right side are more stable in the 2H phase. For binary alloys, the average adsorption energy across two distinct adsorption sites was considered. The optimal adsorption energy, $\Delta \mathit{H}_{\mathrm{ads}}^{\mathrm{H}^*} = -0.25$ \text{eV}\, corresponding to a thermodynamically neutral free energy of binding, \( \Delta G_{\mathrm{H}^*} = 0 \, \text{eV} \),\cite{Nrskov2005TrendsEvolutione} is indicated by the horizontal dashed line in both figures.
}
  \label{fgr:example}
\end{figure}

Figure 2 provides a more detailed picture of the phase-dependent hydrogen binding on the binary alloys. While Figure 1b showed the average adsorption energy over the two distinct binding sites of an ordered binary with a $2\times 2$ periodicity, Figure 2 shows the range of binding energies across the two types of sites. The data clearly shows that increased stability in the 1T (2H) phase, represented by blue (red) bars correlates with enhanced catalytic activity in the opposite phase, as indicated by red (blue) markers. Furthermore, the formation and free energy calculations for binary alloys (see Figure S2) shows, not surprisingly, that an alloy containing two elements each of which are most stable in the same phase, will also prefer that phase.   
 However, alloying elements with different phase preferences can produce more complex behavior. Combinations of group-IV and group-VI TMs, particularly those involving Mo and W, tend to be unstable in both phases. In contrast, alloying group-V elements, which have a slight preference for the 2H phase, with group-IV elements that prefer the 1T phase can stabilize, or nearly stabilize, alloys in the 1T structure.

\begin{figure*}
 \centering
 \includegraphics[height=9.5cm]{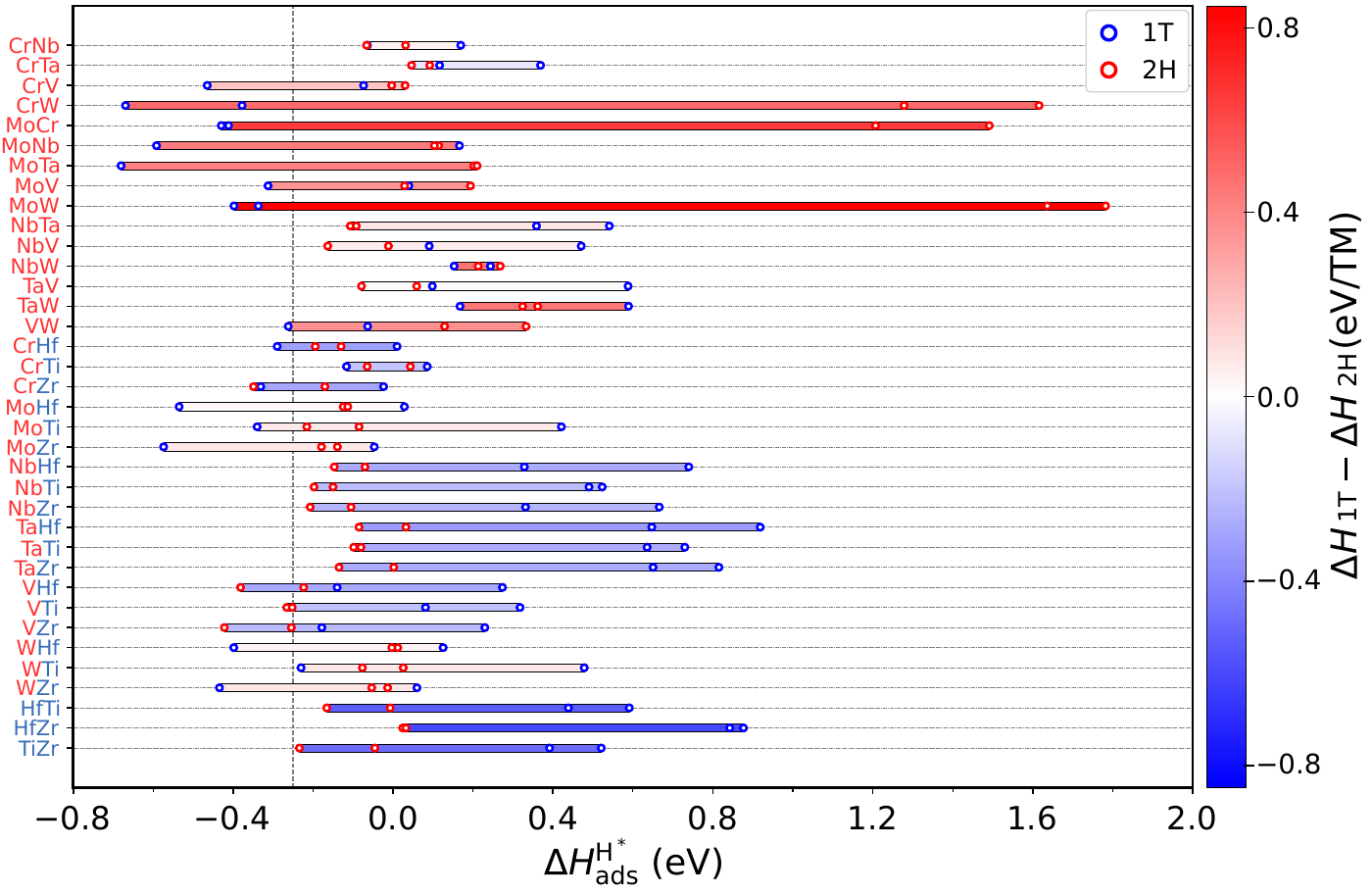}
 \caption{Hydrogen adsorption energy for all possible binary TMDCs with equimolar concentration (AB$\mathrm{S}_2$) in the 1T and 2H phases. The circles represent the hydrogen adsorption energy at two distinct adsorption sites for each structure. Red (blue) circles correspond to adsorption on the 2H (1T) structure. The color of the bars reflects the relative stability of the alloy in the 2H and 1T phase, respectively. Blue bars correspond to structures that are more stable in the 1T phase, whereas red bars indicate those more stable in the 2H phase. As a general trend, hydrogen adsorbs more strongly on the less stable phase.}
 \label{fgr:example2col}
\end{figure*}

The fact that high basal plane activity is predominantly  seen for TMDCs in their unstable phase obviously poses a challenge. Luckily, it turns out that the activity-stability correlation is to a large extent a feature of the \emph{local} atomic environment of the binding site.  
This is illustrated by Figure 3, which shows that the binding energy of hydrogen on the binary alloy (CrZr)S$_2$ is largely determined by the type of TMs occupying the sites closest to the S adsorption site. In other words, H bonding on (CrZr)S$_2$ is similar to CrS$_2$ at S atoms surrounded by Cr and is similar to ZrS$_2$ at S atoms surrounded by Zr. Since ZrS$_2$ is stable in the 1T phase and CrS$_2$ is stable in the 2H phase, Cr atoms (Zr atoms) can be regarded as "locally unstable atoms" in the 1T phase (2H phase) of (CrZr)S$_2$. 
The trend shown in Figure 3 holds for (CrZr)S$_2$ in both the 2H and 1T phase and for other binary alloys as well.


\begin{figure*}
 \centering
 \includegraphics[height=4.5cm]{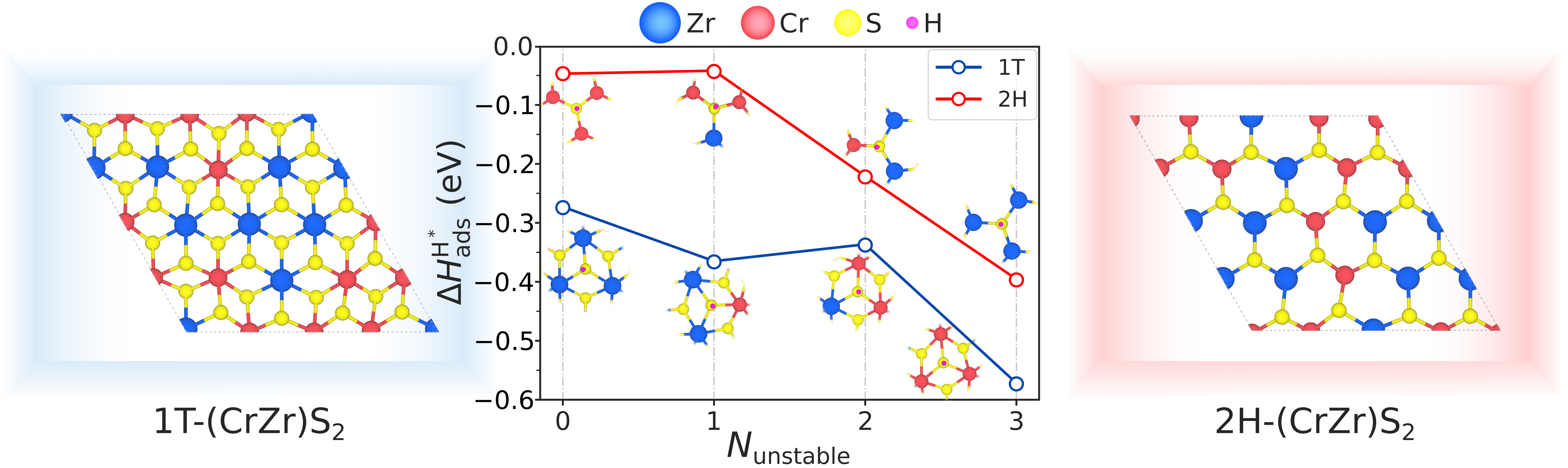}
 \caption{Hydrogen adsorption energy on (CrZr)S$_2$ as a function of the number of locally unstable TM atoms ($N_{\text{unstable}}$). A TM atom is locally unstable if its unary TMDC is more stable in the opposite phase.
 Four distinct adsorption sites were considered for each structure in both the 1T and 2H phases as depicted in the left and right panels, respectively. }
 \label{fgr:example2col}
\end{figure*}

The projected density of states (PDOS) analysis shown in Figure 4 offers insights into the underlying mechanisms of catalytic activity at different sulfur sites. Specifically, the PDOS of a S-$p_z$ orbital in (CrZr)S$_2$ in the 1T-phase (a) and 2H-phase (b) is examined for sulfur atoms in different local environments. The local environments are denoted S0 to S3 corresponding to the number of locally unstable nearest neighbor TM atoms ranging from 0 til 3 (the same binding sites as considered in Figure 3). The center of the $p_z$ band (over both the occupied and unoccupied part) is calculated and indicated in the upper right corner. For both phases we see a correlation between the  $p_z$ band center ($\varepsilon_p$) and the H binding energy shown in Figure 3. For both phases, configurations with higher $\varepsilon_p$, i.e. closer to the Fermi level, exhibit stronger binding of H. This finding aligns with previous studies\cite{Chen2023BasalPrediction}. We conclude that the enhanced catalytic activity at more locally unstable sites, i.e. with more TM nearest neighbor atoms preferring the opposite phase, is due to an increased availability of states with $p_z$ character close to the Fermi level. 

Our findings so far reveal a clear correlation between the local stability and catalytic activity in the TMDCs. The challenge then becomes to create globally stable structures that are locally unstable (or precisely contain TM atoms in their unstable phase). The binary alloys do not seem to provide sufficient flexibility to meet this goal. Despite several of them displaying promising catalytic activity, most of those are thermodynamically unstable (see Figure S2). Additionally,, the low number of different metal atoms restricts the variety of adsorption sites available. Given the complexity of catalytic reactions in general, a diverse set of adsorption sites defined by different local environments is advantageous. To address these limitations, the next step is to systematically increase the number of TM elements.

\begin{figure}[h]
\centering
  \includegraphics[height=6cm]{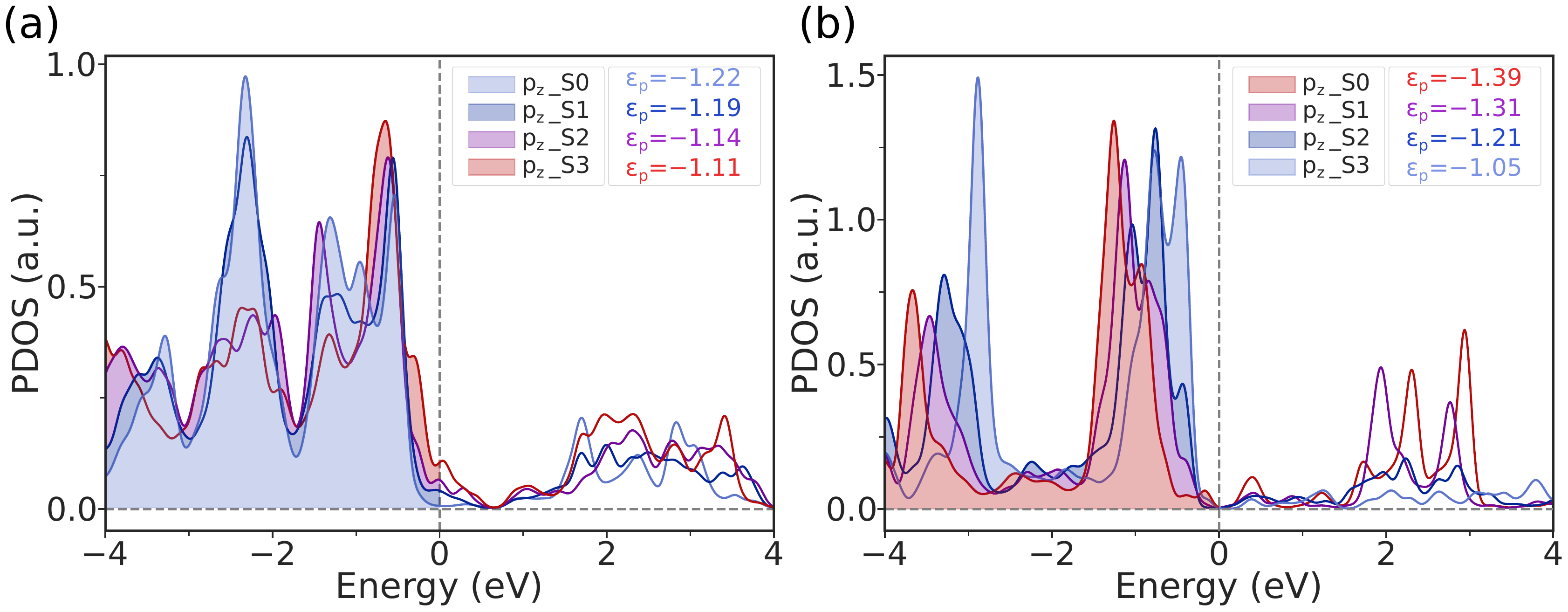}
  \caption{Projected density of states (PDOS) for the p$_z$ orbital of sulfur atoms in the binary TMDC CrZrS$_2$ for both (a) 1T and (b) 2H phases. S0 to S3 represent the specific sulfur sites where hydrogen adsorption will occur, though these figures depict the system without hydrogen. The structures and adsorption sites are shown in Figure 3. Catalytic activity in both configurations follows the trend S3 $>$ S2 $>$ S1 $>$ S0. $\varepsilon_p$ denotes the p-band center relative to the Fermi level, with shaded regions indicating the occupied states.
}
  \label{fgr:example}
\end{figure}

\subsection{Path to High-Entropy Alloy Catalysts}

This section presents a systematic approach for identifying stable HEAs composed of five metal atoms and an activated basal plane for HER. We refer to TM elements stable in the 1T and 2H phase of their unary TMDCs as "T-phase elements" and "H-phase elements", respectively. As mentioned earlier, this investigation focuses on three T-phase and six H-phase elements. Our approach begins by incrementally increasing the number of elements to form ternary, quaternary, and pentanary alloys that are stable in the 1T phase. Once this strategy is validated for the 1T phase, it will be applied directly to the 2H phase. Based on free energy calculations for binary alloys (Figure S2), W was excluded from further investigation, as its combination with any T-phase elements consistently results in highly unstable structures. Therefore, we proceeded with the remaining five H-phase elements.

In our search for stable ternary alloys in the 1T phase we only considered structures containing at least one T-phase element, as alloys based purely on H-phase elements are not likely to be stable. This resulted in 46 possible equimolar ternary combinations, of which 20 were found to be stable in the 1T phase (with the criterion $\Delta G < 0.05$ eV/TM atom), while exhibiting significant instability in the 2H phase. As shown in Figure S3, all H-phase elements can be stabilized in at least one alloy composition, except for Mo, which produces unstable 1T structures in combination with any other elements. For this reason Mo was excluded from further investigation. Hydrogen adsorption energies were calculated for all possible adsorption sites of the stable structures, and the results are presented in Figure S4. Although some stable structures were identified, most remained inactive for HER, underscoring the need to explore quaternary alloys. Further details on the selection of random configurations and hydrogen adsorption sites are discussed in the Methods section.

\begin{figure*}
 \centering
 \includegraphics[height=10cm]{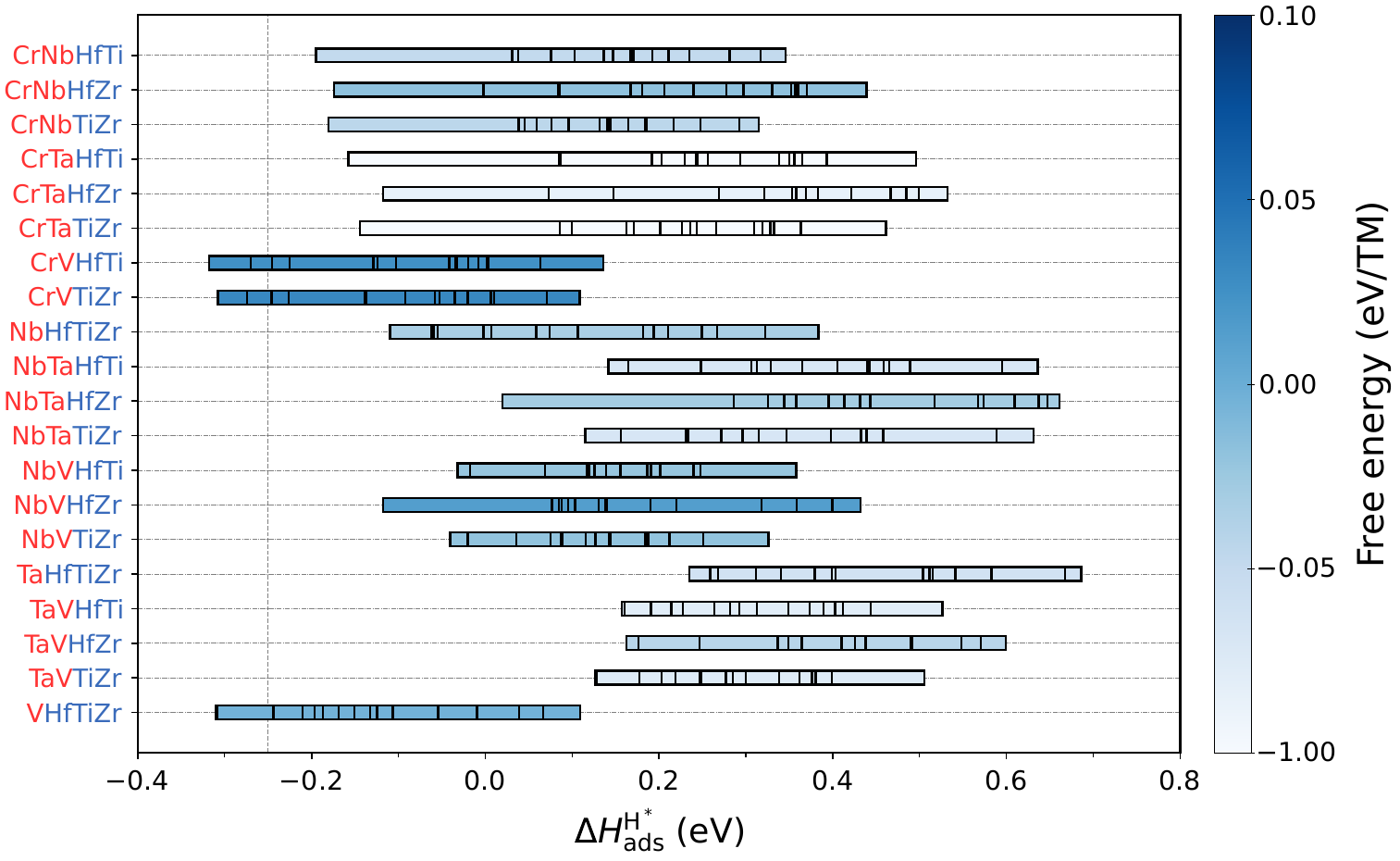}
 \caption{Range of hydrogen adsorption energies on all stable quaternary TMDCs with equimolar concentration (ABCD)$\mathrm{S}_2$ in the 1T phase. Adsorption energy values across different adsorption sites are indicated by vertical black lines, while the colormap represents the free energy of each structure at 1000 K.}
 \label{fgr:example2col}
\end{figure*}

To identify stable quaternary alloys, we modified the selection criteria to include only configurations with at least two T-phase elements. This adjustment led to 22 possible quaternary alloys of which 20 were stable in the 1T phase and highly unstable in the 2H phase (Figure S4). As shown in Figure 5, these quaternary alloys represent an improvement over their ternary counterparts, with several alloys demonstrating hydrogen adsorption energies closer to the optimal value (vertical dashed line). This improvement underscores the enhanced potential of quaternary alloys over ternary alloys. To investigate if the already promising properties of the quaternary alloys can be further optimised we next consider pentanary HEAs containing five TM atoms.

\begin{figure*}
 \centering
 \includegraphics[height=9cm]{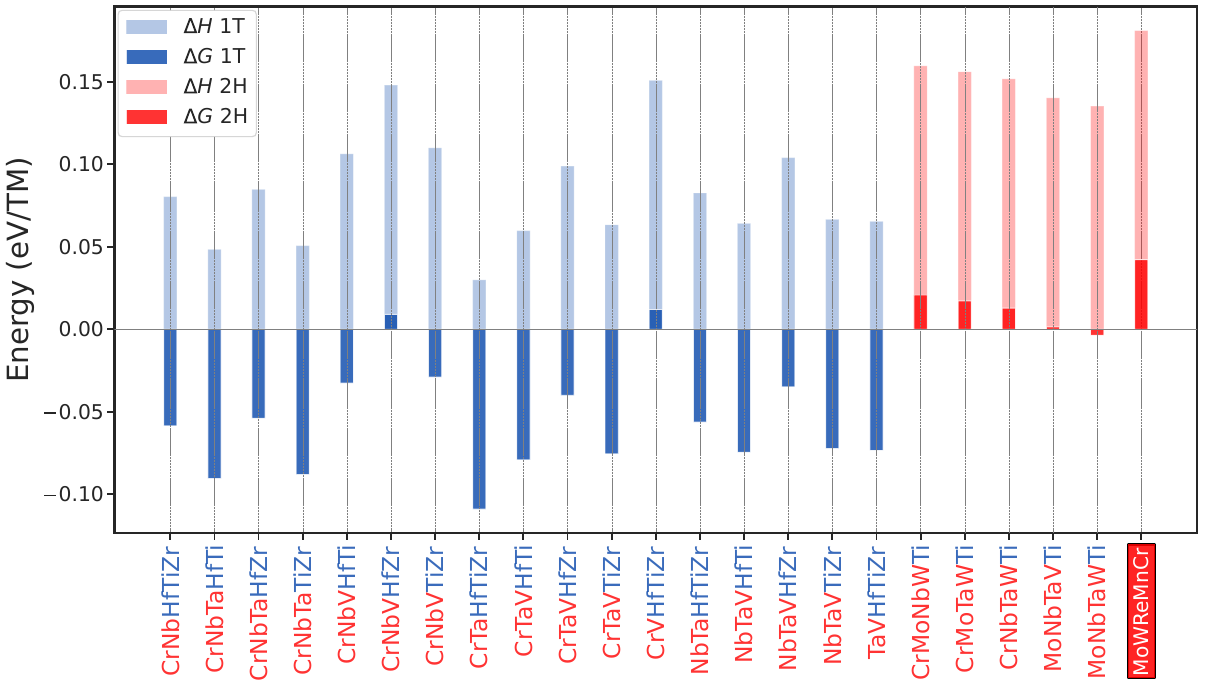}
 \caption{Thermodynamic stability of all stable pentanary TMDCs with equimolar concentration (ABCDE)$\mathrm{S}_2$ in the 1T and 2H phases, depicted by enthalpy of formation and free energy calculations at 1000 K. The experimentally synthesized HEA (MoWReMnCr$\mathrm{S}_2$) is highlighted with a red label.}
 \label{fgr:example2col}
\end{figure*}

Limiting to compositions with at least two T-phase elements and excluding Mo and W, we obtain 17 unique equimolar HEAs in the 1T phase. As shown in Figure 6, all predicted HEAs exhibit good stability properties exhibiting negative free energies, thus validating our alloy exploration strategy. In addition to the 1T structures, all possible compositions of pentanary alloys in the 2H phase containing at most one T-phase element were calculated. The compounds satisfying $\Delta G < 0.05$ eV/TM atom are shown by the red bars in Figure 6. For comparison, the experimentally synthesized HEA (MoWReMnCr)$\mathrm{S}_2$, recently reported as a promising catalyst for HER\cite{Qu2023AElectrocatalysis}, is also depicted in the last column. Although the experimental structure is not equimolar, we replicated the exact stoichiometry from the synthesis, and the resulting stability and catalytic activity were highly consistent with the equimolar structure shown in Figure 6. The experimentally synthesised HEA has higher free energy than the 22 pentanary HEAs predicted in our study. This strongly suggests that the predicted HEAs are indeed synthesizable under the right experimental conditions. 

To assess the phase stability of the predicted HEAs, we evaluated the free energy in the alternate phase. The result shows that these alloys are all highly unstable in the opposite phase, with enthalpy of formation exceeding those of their stable phases by more than 0.3 eV/TM atom. Such a substantial enthalpy difference between the stable and unstable phases indicates a strong thermodynamic preference for the predicted phase. This phase stability gap is a crucial factor in practical synthesis, as it reduces the likelihood of phase mixing or unintended formation of metastable phases during experimental preparation. Further supporting thermodynamic stability, the energy above the convex hull calculations shows that all predicted HEAs have $\Delta \mathit{H}_{\mathrm{hull}}$ values below 0.2 eV/TM atom, indicating that these alloys are close to the most stable combination of their constituent elements and are unlikely to decompose into other phases. The calculated values of enthalpy of formation, free energy of mixing, and energy above the convex hull for the predicted HEAs can be found in Table S1.

To evaluate the catalytic potential of the predicted HEAs, hydrogen adsorption energies were calculated across all possible adsorption sites. As the pentanary alloys were represented in a $5\times 5$ supercell, a total of 25 distinct adsorption energies were evaluated per alloy. As illustrated in Figure 7, the predicted HEAs demonstrate high catalytic activity, with a narrow range of adsorption energies close to the optimal value. Moreover, it is evident that the alloys predicted in this study exhibit not only superior stability but also higher catalytic activity compared to the previously reported experimental structure.

\begin{figure*}
 \centering
 \includegraphics[height=16cm]{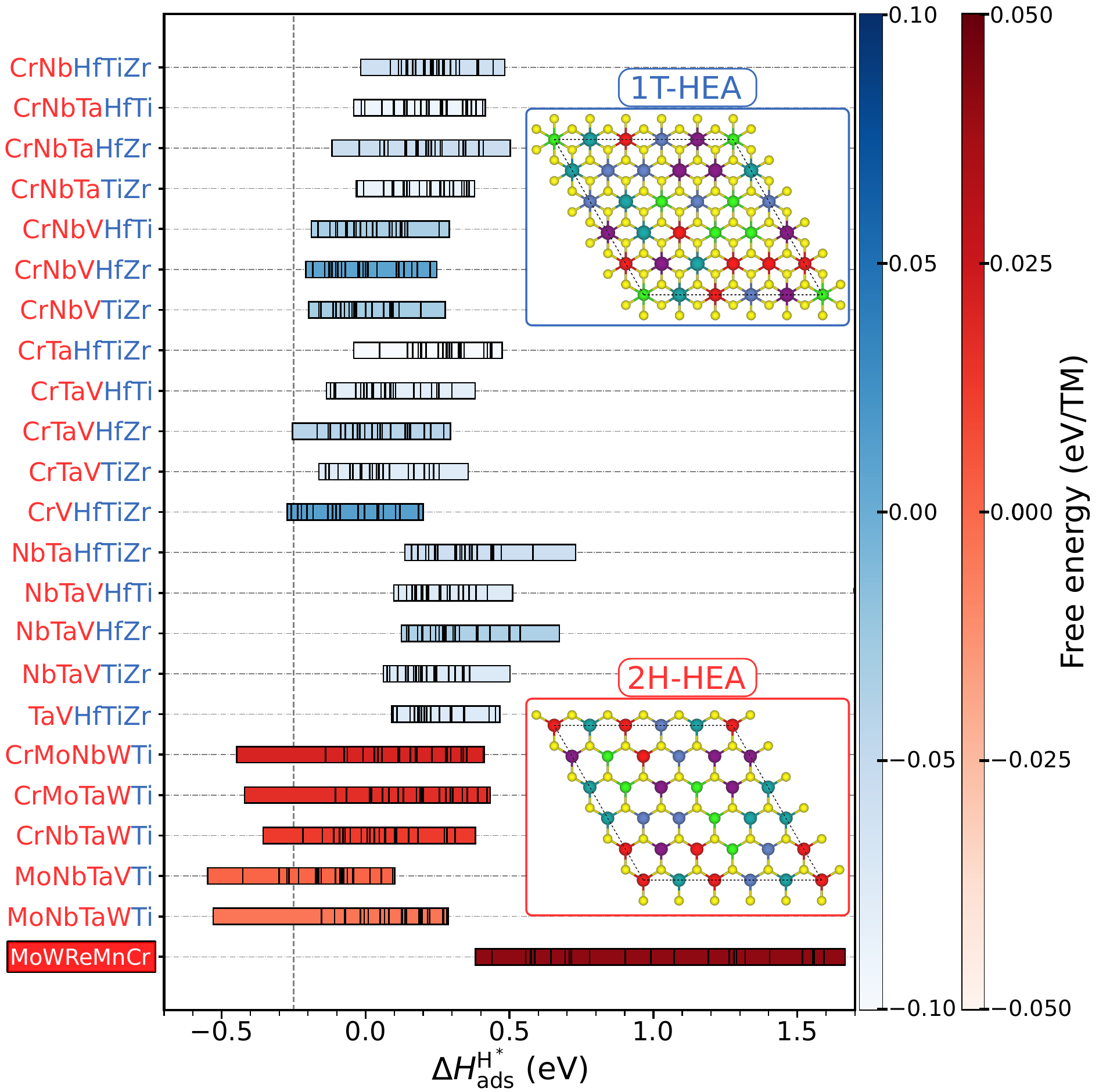}
 \caption{Range of hydrogen adsorption energies on all stable pentanary TMDCs with equimolar concentration (ABCDE)$\mathrm{S}_2$ in the 1T and 2H phases. Adsorption energy values at different adsorption sites are represented by vertical black lines, while the blue and red colormaps indicate the free energy of the 1T and 2H structures at 1000 K, respectively. The experimentally synthesized HEA (MoWReMnCr)$\mathrm{S}_2$ is highlighted with a red label for comparison.
}
 \label{fgr:example2col}
\end{figure*}

\section{Conclusions}
Based on density functional theory calculations we have predicted a number of transition metal dichalcogenide (TMDC) high entropy alloys (HEAs) with good thermodynamic stability, as quantified by low free energies of formation, and high catalytic activity towards hydrogen evolution on the basal plane. We have established a clear correlation between the local structural stability and activity of a catalytic site. In particular, we found that the transition metals that prefer to crystallise in the 1T-phase can enhance the catalytic activity when alloyed into a 2H-phase TMDC, and vice versa. We further showed that it is possible to form TMDC alloys containing both 1T and 2H-phase elements that are stabilised by the configurational entropy. The resulting TMDC HEAs with mixed 1T/2H composition where shown to possess lower free energy of formation and higher catalytic activity compared to a previously synthesised 2H-TMDC HEA. Our work provides insight into the fundamental relation between local structural stability and catalytic activity and proposes TMDC HEAs as a promising class of non-precious hydrogen evolution catalysts.

\begin{acknowledgement}
K. S. T. is a Villum Investigator supported by VILLUM FONDEN (grant no. 37789).

\end{acknowledgement}

\begin{suppinfo}

Information about the choice of exchange-correlation functional, enthalpy of formation, free energy of mixing, hydrogen adsorption energy, and energy above the convex hull for various TMDC alloys, including binary, ternary, quaternary, and pentanary configurations, with atomic structures referenced in the article, as well as different random configurations and SQSs.

\end{suppinfo}

\bibliography{achemso-demo}

\end{document}





\renewcommand{\thefigure}{S\arabic{figure}} %
\renewcommand{\thetable}{S\arabic{table}} %

\begin{figure*}
 \centering
 \includegraphics[height=8.5cm]{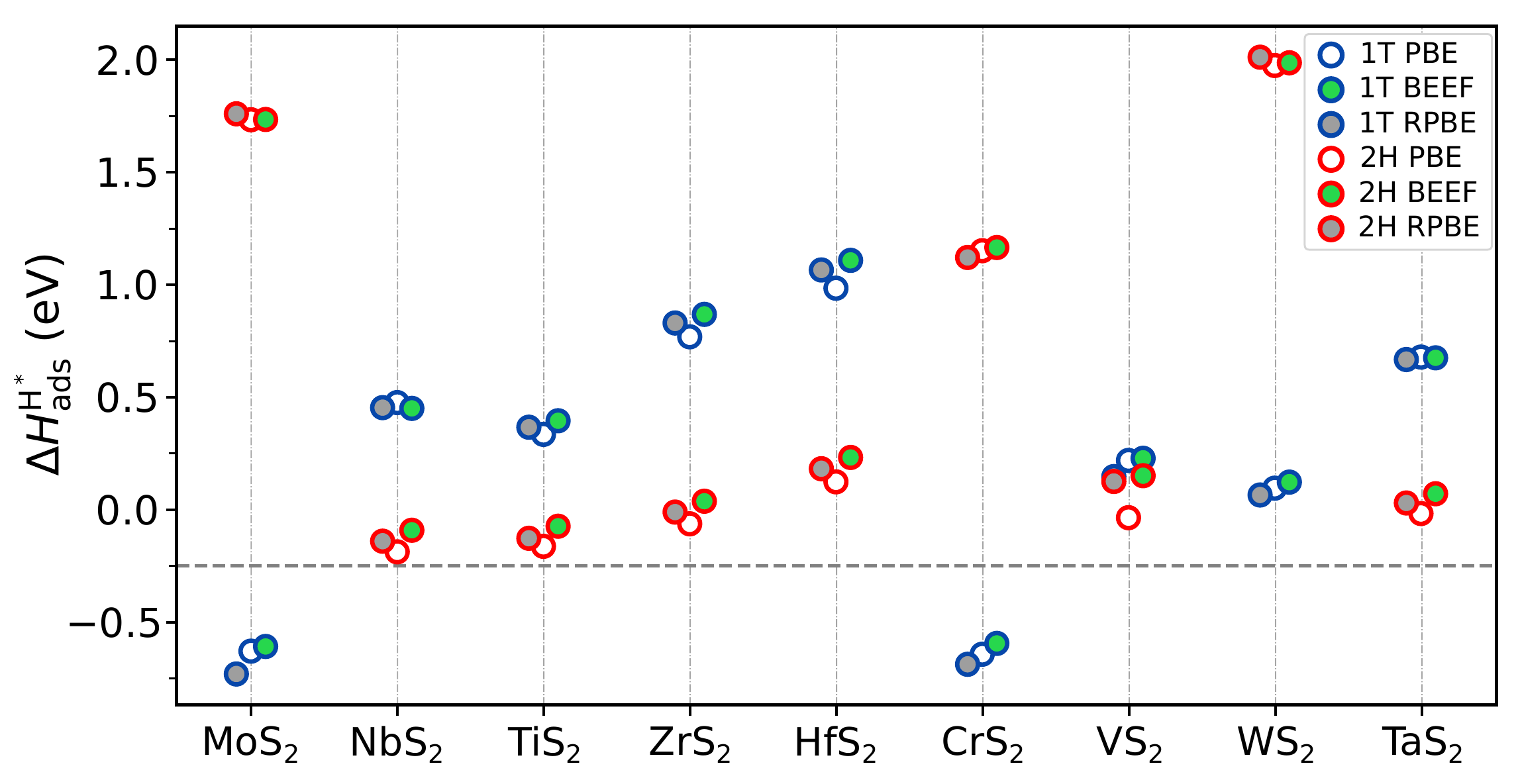}
 \caption{Hydrogen adsorption energies on unary TMDCs, calculated using various exchange-correlation functionals: PBE, BEEF-vdW, and RPBE, for both the 1T and 2H phases.}
 \label{fgr_functionals}
\end{figure*}

\begin{figure*}
 \centering
 \includegraphics[height=8.5cm]{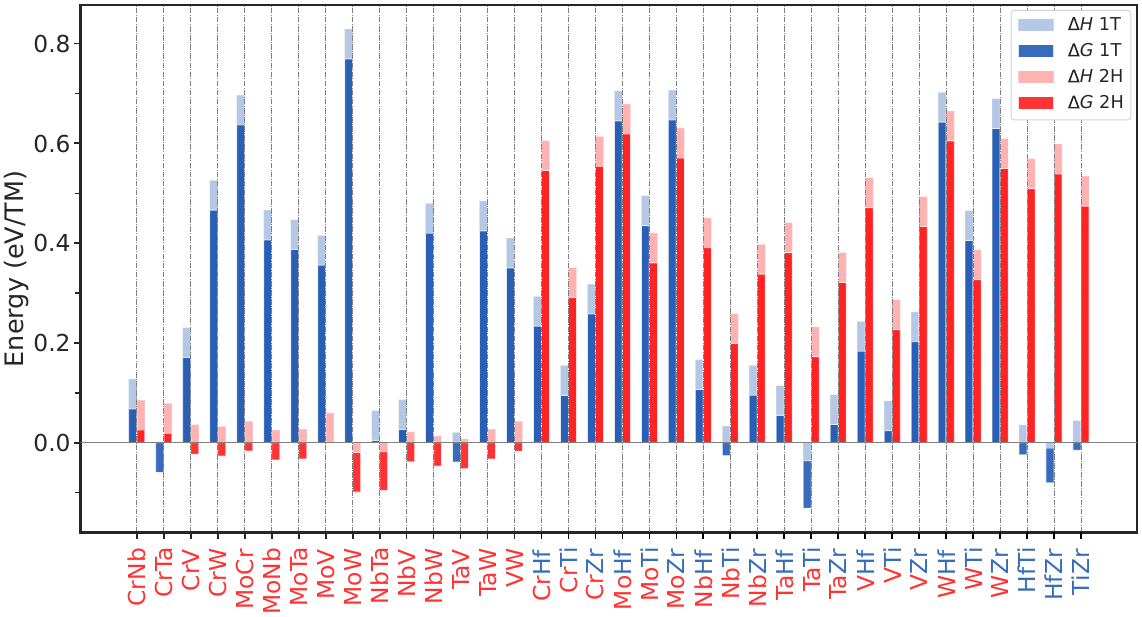}
 \caption{Thermodynamic stability of all possible combinations of binary TMDCs with equimolar concentrations (AB)$\mathrm{S}_2$ in the 1T and 2H phases, represented through enthalpy of formation and free energy calculations at 1000 K. The red/blue color scheme applied to the labels indicates the most stable phase for each element in its unary structure, with red representing the 2H phase and blue representing the 1T phase.}
 \label{fgr:example2col}
\end{figure*}

\begin{figure*}
 \centering
 \includegraphics[height=8.5cm]{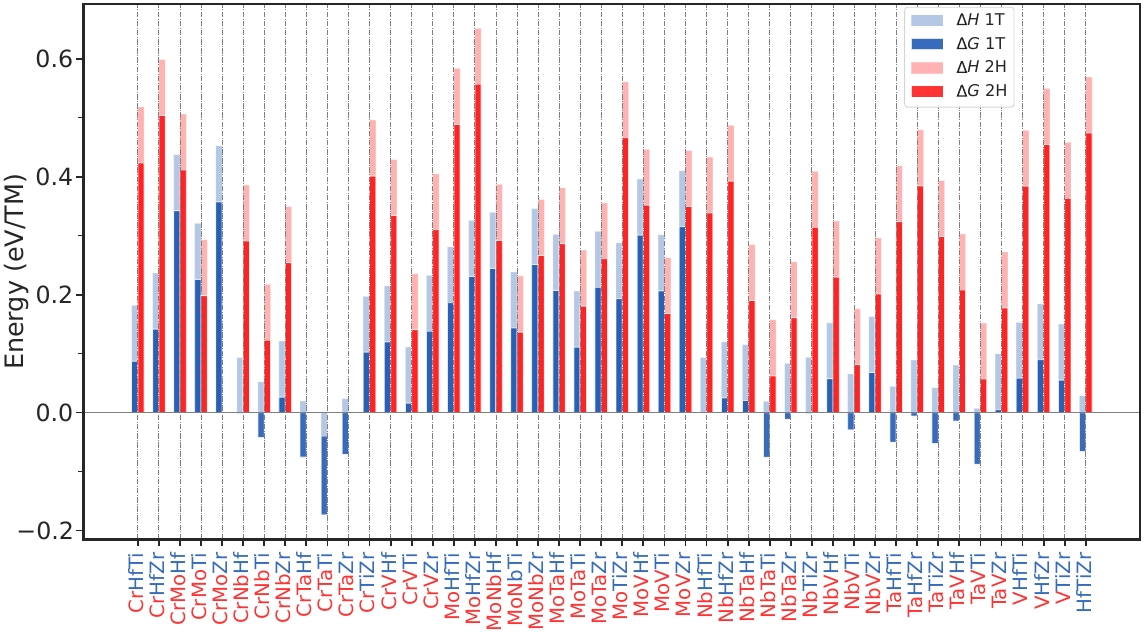}
 \caption{Thermodynamic stability of all possible combinations of ternary TMDCs with equimolar concentrations (ABC)$\mathrm{S}_2$ in the 1T and 2H phases, represented through enthalpy of formation and free energy calculations at 1000 K. The red/blue color scheme applied to the labels indicates the most stable phase for each element in its unary structure, with red representing the 2H phase and blue representing the 1T phase.}
 \label{fgr:example2col}
\end{figure*}

\begin{figure*}
 \centering
 \includegraphics[height=10cm]{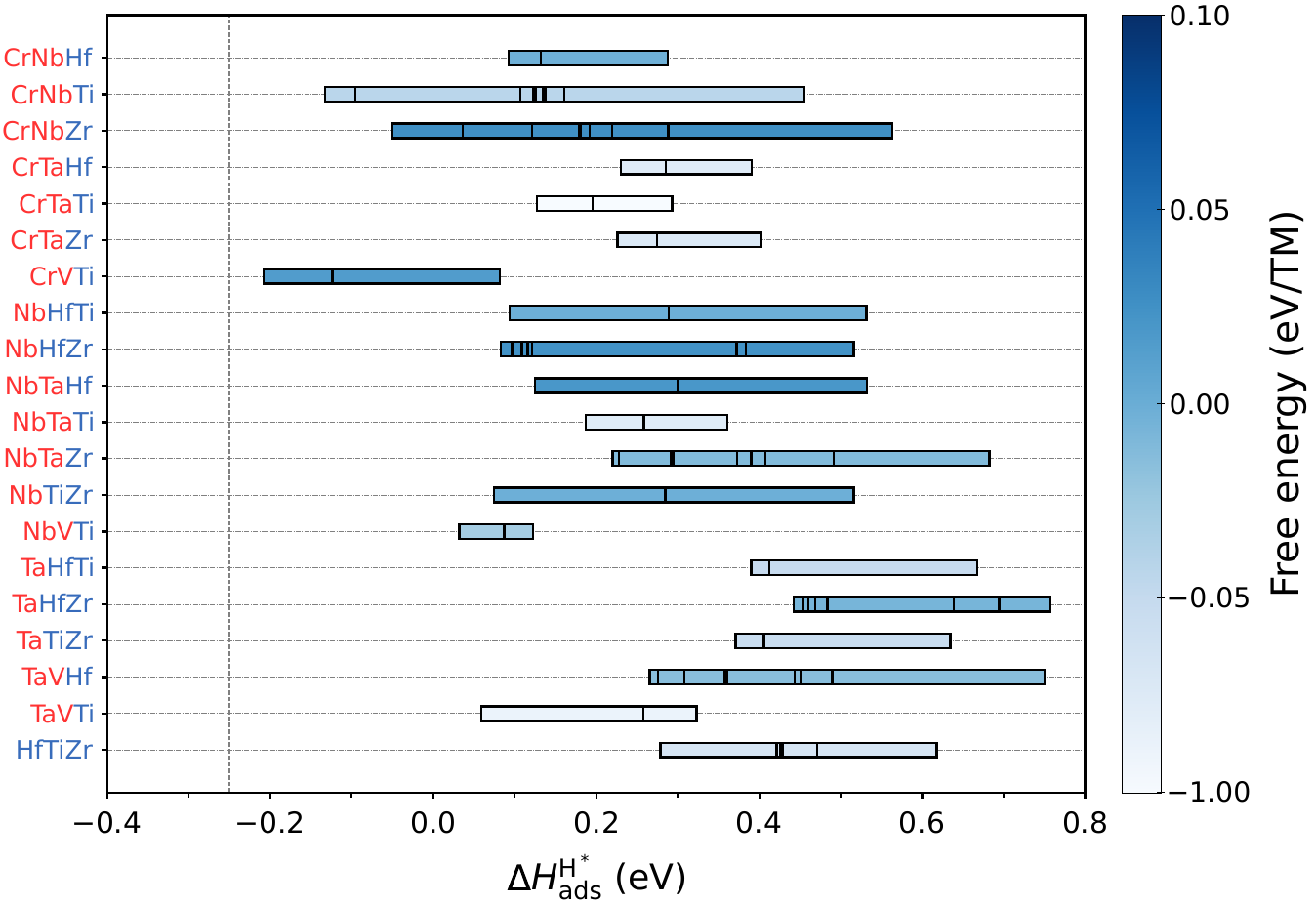}
 \caption{Range of hydrogen adsorption energies on all stable ternary TMDCs with equimolar concentration (ABC)$\mathrm{S}_2$ in the 1T phase. Adsorption energy values for different adsorption sites are indicated by vertical black lines, while the colormap represents the free energy of each structure at 1000 K.}
 \label{fgr:example2col}
\end{figure*}

\begin{figure*}
 \centering
 \includegraphics[height=8.5cm]{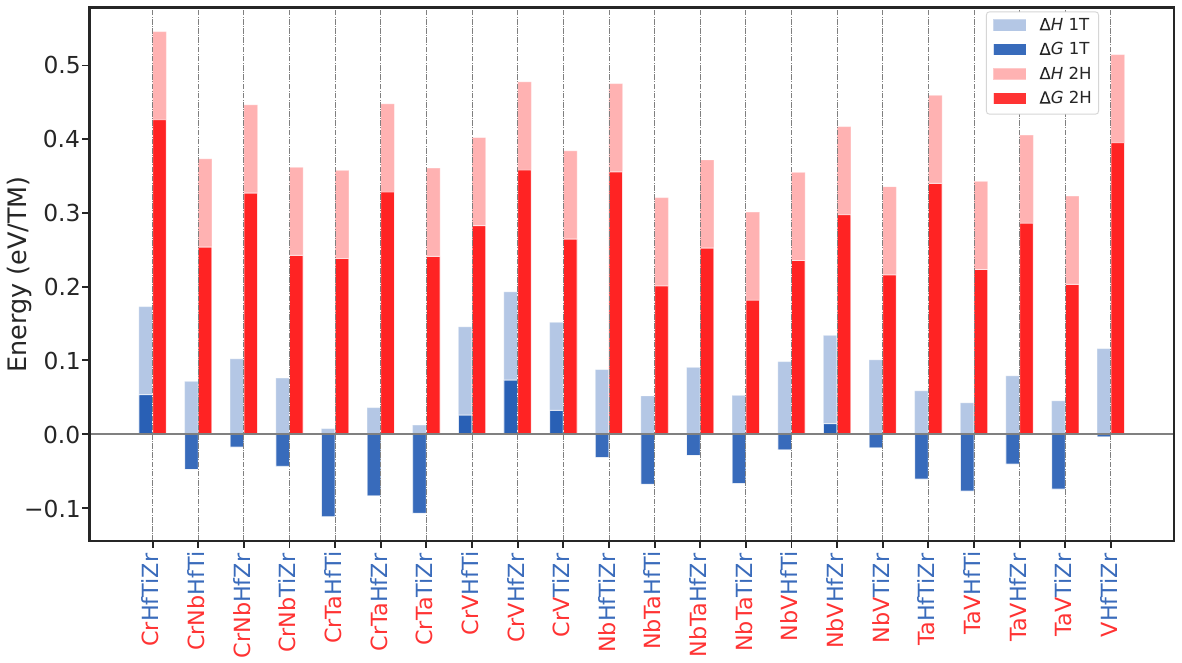}
 \caption{Thermodynamic stability of quaternary TMDCs with equimolar concentrations (ABCD)$\mathrm{S}_2$ in the 1T and 2H phases, represented through enthalpy of formation and free energy calculations at 1000 K. The red/blue color scheme applied to the labels indicates the most stable phase for each element in its unary structure, with red representing the 2H phase and blue representing the 1T phase.}
 \label{fgr:example2col}
\end{figure*}

\begin{figure*}
 \centering
 \includegraphics[height=7cm]{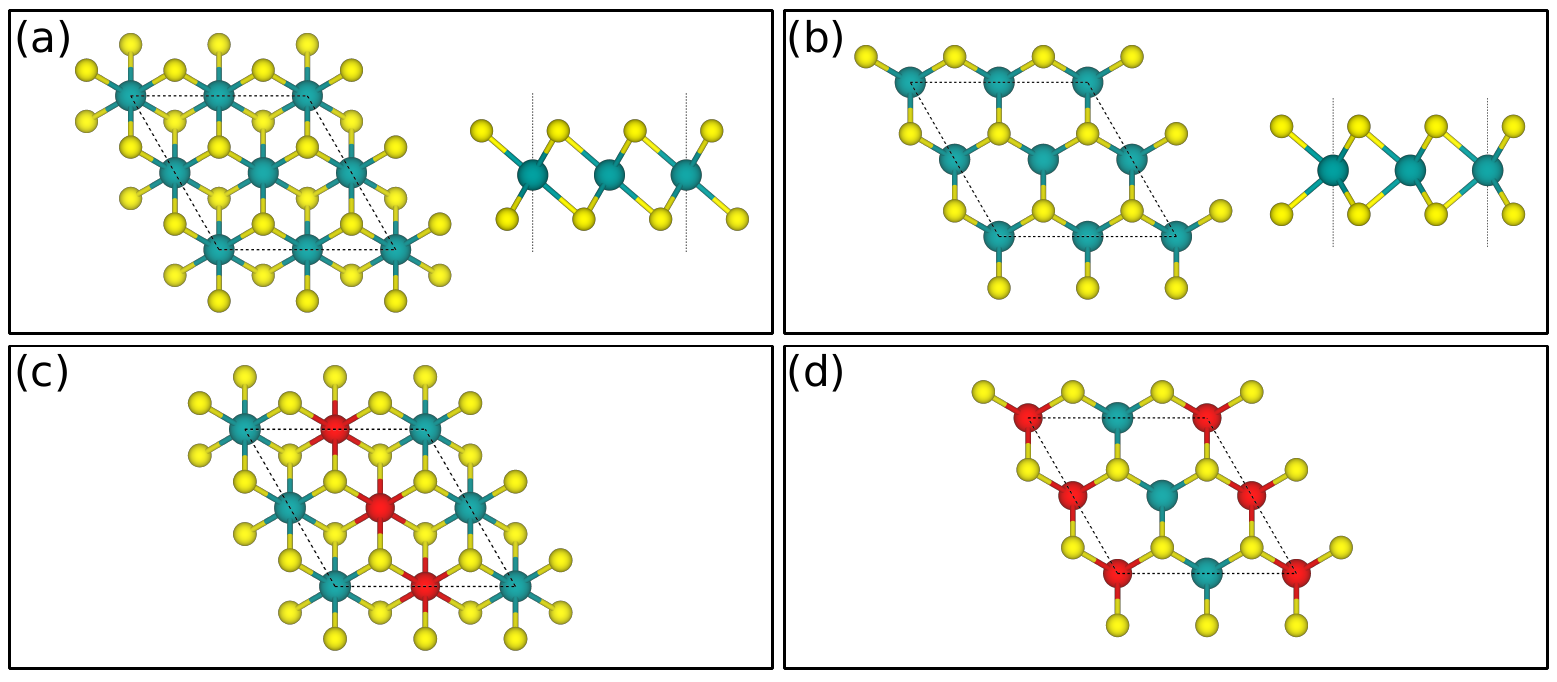}
 \caption{Top and side views of 2$\times$2 unary TMDCs in the (a) 1T and (b) 2H phases, included in all calculations. Top views of 2$\times$2 binary TMDCs with equimolar concentration (AB)$\mathrm{S}_2$ in the (c) 1T and (d) 2H phases, as considered in all computational analyses.}
 \label{fgr:example2col}
\end{figure*}

\begin{figure*}
 \centering
 \includegraphics[height=13.5cm]{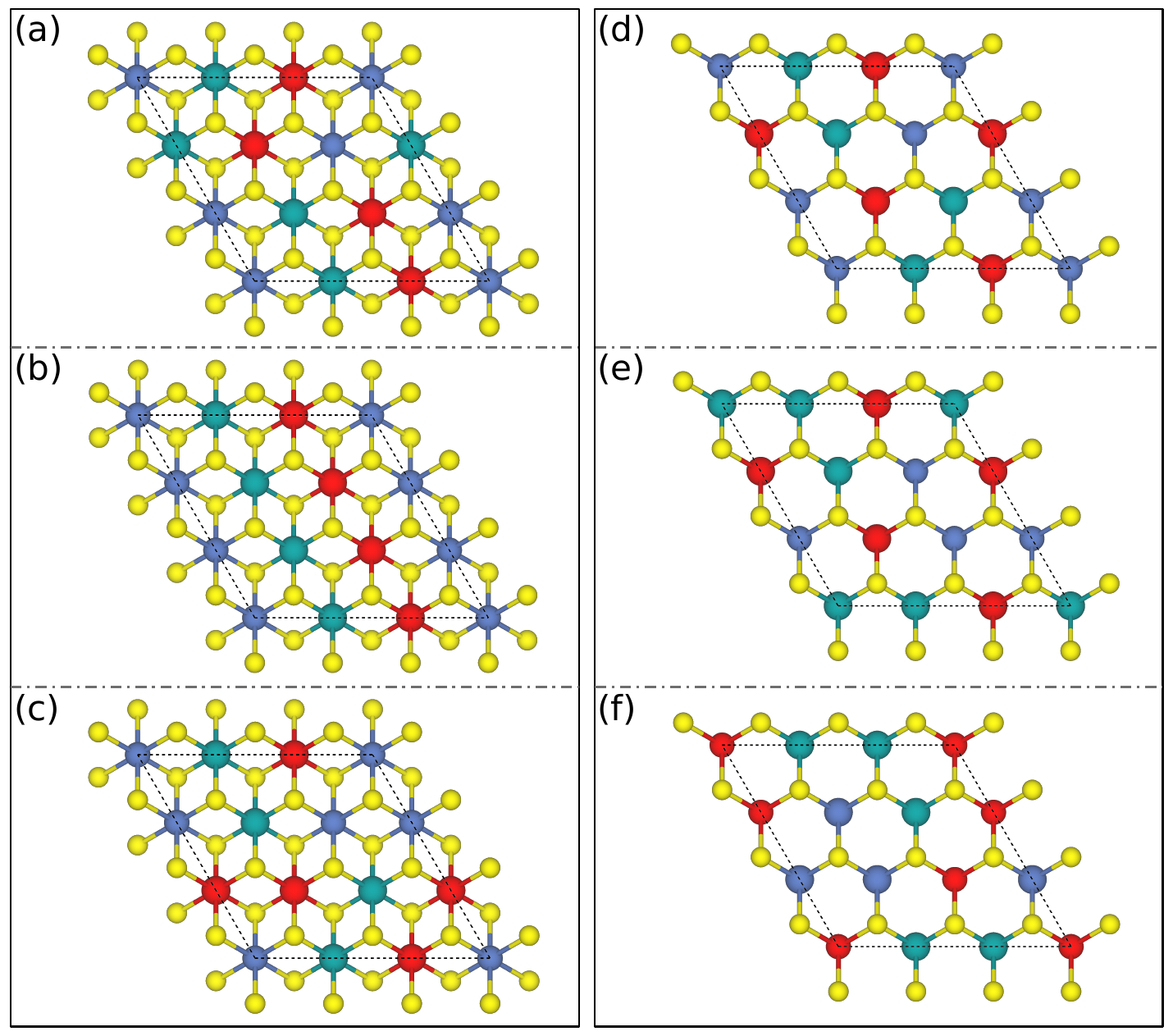}
 \caption{Different random configurations of 3$\times$3 ternary TMDCs with equimolar concentration (ABC)$\mathrm{S}_2$ in the (a-c) 1T and (d-f) 2H phases, as considered in all calculations.}
 \label{fgr:example2col}
\end{figure*}

\begin{figure*}
 \centering
 \includegraphics[height=13.5cm]{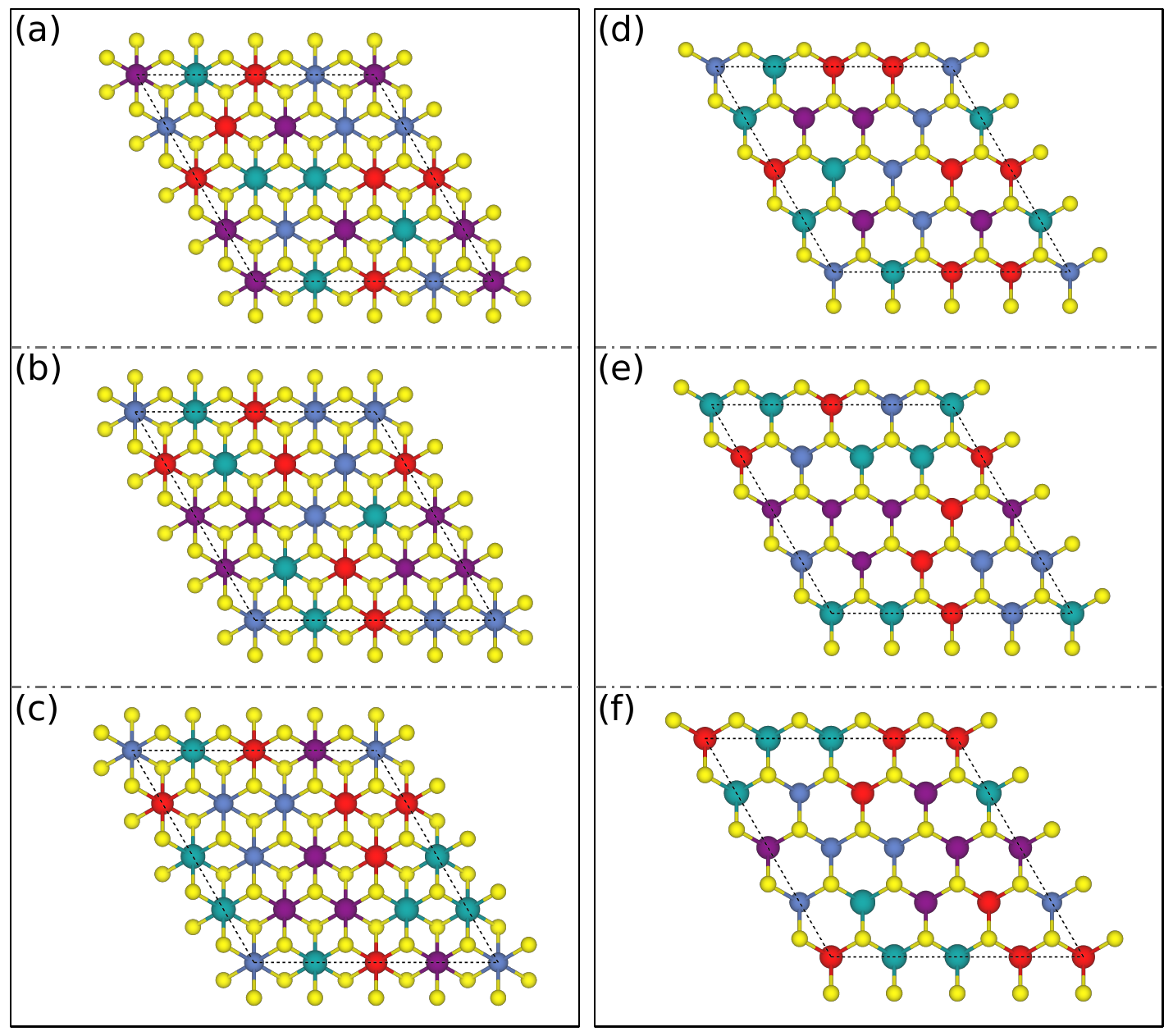}
 \caption{Different random configurations of 4$\times$4 quaternary TMDCs with equimolar concentration (ABCD)$\mathrm{S}_2$ in the (a-c) 1T and (d-f) 2H phases, as considered in all calculations.}
 \label{fgr:example2col}
\end{figure*}

\begin{figure*}
 \centering
 \includegraphics[height=5.3cm]{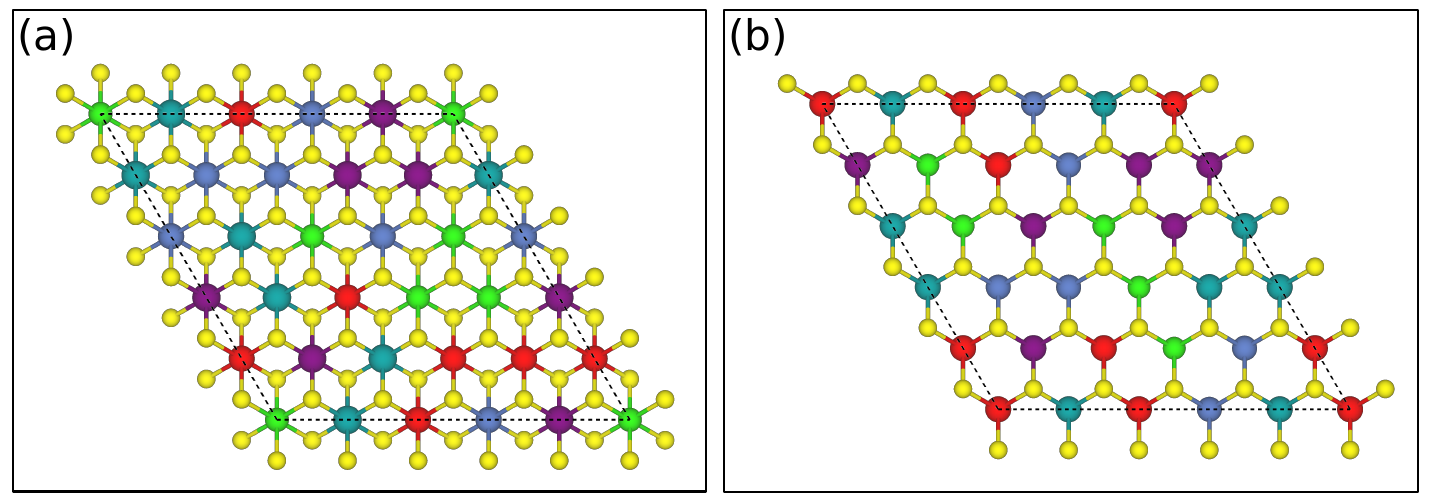}
 \caption{Special quasi-random structures (SQS) of 5$\times$5 pentanary TMDCs with equimolar concentration (ABCDE)$\mathrm{S}_2$ in the (a) 1T and (b) 2H phases, as considered in all calculations.}
 \label{fgr:example2col}
\end{figure*}

\begin{table}[ht]
    \centering
    \caption{Enthalpy of formation (\( \Delta H_f \)), free energy of mixing (\( \Delta G_{\mathrm{mix}} \)) at 1000 K, and energy above the convex hull ($\Delta \mathit{H}_{\mathrm{hull}}$) for all high-entropy alloys at their more stable phase.}
    \renewcommand{\arraystretch}{1.5}
    \begin{tabular}{c c c c c}
        \toprule
        \rowcolor{gray!20}
        \textbf{Alloy Composition} &\textbf{Phase} &\textbf{\( \Delta H_f \) (eV/TM)} & \textbf{\( \Delta G_{\mathrm{mix}} \) (eV/TM)} & \textbf{$\Delta \mathit{H}_{\mathrm{hull}}$ (eV/TM)} \\
        \midrule
        \centering (CrNbHfTiZr)$_\text{0.2}$S$_\text{2}$ & 1T & 0.0804 & -0.0585 & 0.1193 \\
        \rowcolor{gray!10}
        \centering (CrNbTaHfTi)$_\text{0.2}$S$_\text{2}$ & 1T & 0.0484 & -0.0905 & 0.0817 \\
        \centering (CrNbTaHfZr)$_\text{0.2}$S$_\text{2}$ & 1T & 0.0848 & -0.0541 & 0.1218 \\
        \rowcolor{gray!10}
        \centering (CrNbTaTiZr)$_\text{0.2}$S$_\text{2}$ & 1T & 0.0507 & -0.0882 & 0.1028 \\
        \centering (CrNbVHfTi)$_\text{0.2}$S$_\text{2}$ & 1T & 0.1062 & -0.0327 & 0.1271 \\
        \rowcolor{gray!10}
        \centering (CrNbVHfZr)$_\text{0.2}$S$_\text{2}$ & 1T & 0.1479 & 0.0090 & 0.1851 \\
        \centering (CrNbVTiZr)$_\text{0.2}$S$_\text{2}$ & 1T & 0.1099 & -0.0290 & 0.1487 \\
        \rowcolor{gray!10}
        \centering (CrTaHfTiZr)$_\text{0.2}$S$_\text{2}$ & 1T & 0.0299 & -0.1090 & 0.0583 \\
        \centering (CrTaVHfTi)$_\text{0.2}$S$_\text{2}$ & 1T & 0.0598 & -0.0791 & 0.0883 \\
        \rowcolor{gray!10}
        \centering (CrTaVHfZr)$_\text{0.2}$S$_\text{2}$ & 1T & 0.0989 & -0.0400 & 0.1102 \\
        \centering (CrTaVTiZr)$_\text{0.2}$S$_\text{2}$ & 1T & 0.0633 & -0.0756 & 0.0916 \\
        \rowcolor{gray!10}
        \centering (CrVHfTiZr)$_\text{0.2}$S$_\text{2}$ & 1T & 0.1508 & 0.0119 & 0.1641 \\
        \centering (NbTaHfTiZr)$_\text{0.2}$S$_\text{2}$ & 1T & 0.0826 & -0.0563 & 0.1040 \\
        \rowcolor{gray!10}
        \centering (NbTaVHfTi)$_\text{0.2}$S$_\text{2}$ & 1T & 0.0642 & -0.0747 & 0.0858 \\
        \centering (NbTaVHfZr)$_\text{0.2}$S$_\text{2}$ & 1T & 0.1040 & -0.0349 & 0.1368 \\
        \rowcolor{gray!10}
        \centering (NbTaVTiZr)$_\text{0.2}$S$_\text{2}$ & 1T & 0.0665 & -0.0724 & 0.0973 \\
        \centering (TaVHfTiZr)$_\text{0.2}$S$_\text{2}$ & 1T & 0.0654 & -0.0735 & 0.0868 \\
        \rowcolor{gray!10}
        \centering (CrMoNbWTi)$_\text{0.2}$S$_\text{2}$ & 2H & 0.1597 & 0.0208 & 0.1786 \\
        \centering (CrMoTaWTi)$_\text{0.2}$S$_\text{2}$ & 2H & 0.1561 & 0.0172 & 0.1875 \\
        \rowcolor{gray!10}
        \centering (CrNbTaWTi)$_\text{0.2}$S$_\text{2}$ & 2H & 0.1517 & 0.0128 & 0.1840 \\
        \centering (MoNbTaVTi)$_\text{0.2}$S$_\text{2}$ & 2H & 0.1403 & 0.0014 & 0.1485 \\
        \rowcolor{gray!10}
        \centering (MoNbTaWTi)$_\text{0.2}$S$_\text{2}$ & 2H & 0.1352 & -0.0037 & 0.1572 \\
        \bottomrule
    \end{tabular}
\end{table}
